\documentclass[runningheads]{llncs}

\usepackage{amsmath}
\usepackage[all]{xy}
\usepackage{latexsym}
\usepackage{url}

\newcommand{\vit}[1]{\marginpar{\sc vit}\textbf{#1}}

\newcommand{\ignore}[1]{}

\title{Solving Partial Order Constraints\\
  for LPO Termination}

\author{Michael Codish\inst{1}\thanks{Research performed at the
    University of Melbourne} \and Vitaly Lagoon\inst{2} \and Peter J.
  Stuckey\inst{2}$^,$\inst{3}}

\institute{ Department of Computer Science, Ben-Gurion University,
  Israel \and
  Department of Computer Science and Software Engineering\\
  The University of Melbourne, Australia\\
  \and
  NICTA Victoria Laboratory \\
  \email{mcodish@cs.bgu.ac.il},~~ \email{\{lagoon,pjs\}@cs.mu.oz.au} }

\newcommand{\cA}{{\cal A}}
\newcommand{\cF}{{\cal F}}
\newcommand{\cR}{{\cal R}}
\newcommand{\cT}{{\cal T}}
\newcommand{\cV}{{\cal V}}
\newcommand\tuple[1]{\langle #1 \rangle}
\newcommand\encode[1]{\| #1 \|}

\newcommand{\sset}[2]{\left\{~#1  \left|
      \begin{array}{l}#2\end{array}
    \right.     \right\}}
\newcommand{\set}[1]{\left\{
      \begin{array}{l}#1\end{array}
        \right\}}
\newenvironment{SProg}
     {\begin{small}\begin{tt}\begin{tabular}[c]{l}}%
     {\end{tabular}\end{tt}\end{small}}

\def\ll{\mathbf{[\![}}
\def\rr{\mathbf{]\!]}}
\def\prop#1{\ll \mathrm{#1}\rr}

\newcommand{\xyo}[1]{*=<1pc>[o][F-]{#1}}

\begin{document}
\maketitle

\begin{abstract}

  \begin{sloppypar}
    This paper introduces a new kind of propositional \mbox{encoding}
    for reasoning about partial orders. The symbols in an unspecified
    partial order are viewed as variables which take integer values and
    are interpreted as indices in the order.
    For a partial order statement on $n$ symbols each index is
    represented in $\lceil\log_2 n\rceil$ propositional variables and
    partial order constraints between symbols are modeled on the bit
    representations.
    We illustrate the application of our approach to determine LPO
    termination for term rewrite systems. Experimental results are
    unequivocal, indicating orders of magnitude speedups in comparison
    with current implementations for LPO termination.
    The proposed encoding is general and relevant to other applications
    which involve propositional reasoning about partial orders.
  \end{sloppypar}
\end{abstract}

\section{Introduction}

This paper formalizes a propositional logic over partial orders.
Formul\ae\/ in this logic are just like usual propositional
formul\ae\/ except that propositions are statements about a partial
order on a finite set of symbols. For example,
$(f=g)\land((f>h)\lor(h>g))$ is a formula in this logic. We refer to
the formul\ae\/ of this logic as \emph{partial order constraints}.
There are many applications in computer science which involve
reasoning about (the satisfiability of) partial order constraints.
For example, in the contexts of termination analysis, theorem proving,
and planning. The main contribution of this paper is a new kind of
propositional encoding of partial order constraints in propositional
logic.

Contemporary propositional encodings, such as the one considered in
\cite{KK04}, model the atoms (primitive order relations such as $f=g$
or $f>h$ on symbols) in a partial order constraint as propositional
variables. Then, propositional statements are added to encode the
axioms of partial orders which the atoms are subject to.
For a partial order constraint on $n$ symbols, such encodings typically
introduce $O(n^2)$ propositional variables and involve $O(n^3)$
clauses to express the axioms.
In contrast we propose to model the symbols in a partial order
constraint as integer values (in binary representation). For $n$
symbols this requires $k=\lceil\log_2 n\rceil$ propositional variables
for each symbol. The integer value of a symbol reflects its index in
the partial order. Constraints of the form $(f=g)$ or $(f>h)$ are then
straightforward to encode in $k$-bit arithmetic.

We focus on the application to termination analysis for term rewrite
systems (for a survey see \cite{DJ90}) and in particular on LPO
termination \cite{KL80,Der87}.
Experimental results are unequivocal, surpassing the performance of
current termination analyzers such as TTT \cite{HirokawaM05,ttt} and
AProVe \cite{GieslTSF04,aprove} (configured for LPO).
The underlying approach is directly applicable to more powerful
termination proving techniques, such as those based on dependency
pairs \cite{ArtsG00}, which basically involve the same kind of
constraint solving.

Sections~\ref{sec:semantics} and~\ref{sec:lpo} introduce partial order
constraints and the LPO termination problem.
Section~\ref{sec:Pencode} describes the $O(n^2)$ atom-based
propositional encoding of partial order formul\ae\/ introduced in
\cite{KK04} for LPO termination. Section~\ref{sec:Sencode} introduces
our $O(n\log n)$ symbol-based
encoding. Section~\ref{sec:implementation} describes and evaluates our
implementation for LPO termination which is based on the application
of a state-of-the-art propositional SAT
solver~\cite{minisat-page}. Finally, we present related work and
conclusions.

\section{Partial order constraints}\label{sec:semantics}

Informally, a partial order constraint is just like a formula in
propositional logic except that propositions are atoms of the form
$(f>g)$ or $(f=g)$. The semantics of a partial order constraint is a
set of models. A model is an assignment of truth values to atoms which
is required to satisfy both parts of the formula: the ``propositional
part'' and the ``partial order part''.

\paragraph{\bf Syntax:~}
Let $\cF$ be finite non-empty set of symbols and $\cR=\set{>,=}$
consist of two binary relation symbols on $\cF$.  Since $\cR$ is fixed
we denote by $Atom_\cF$ the set of atoms of the form $(f ~R~ g)$ where
$R \in \cR$ and $f,g\in \cF$.
A partial order constraint on $\cF$ is a propositional formula in
which the propositions are elements of $Atom_\cF$. We sometimes write
$(f\geq g)$ as shorthand for $(f>g) \lor (f=g)$. We denote the set of
atoms occurring in a partial order constraint $\varphi$ by
$Atom(\varphi)$.

\paragraph{\bf Semantics:~}

The symbols in $\cR$ are interpreted respectively as a strict partial
order and as equality (both on $\cF$). Let $\varphi$ be a partial
order constraint on $\cF$. The semantics of $\varphi$ is a set of models.
Intuitively, a model of $\varphi$ is a set of atoms from $Atom_\cF$
which satisfies both parts of the formula: the propositional part
and the partial order part. Before presenting a formal definition we
illustrate this intuition by example.

\begin{example}  \label{ex:syntax}
Let $\cF=\{f,g,h\}$.
The following are partial order constraints:
\begin{eqnarray*}
   \varphi_1 &=& (f>g) \land ((f>h) \lor (h>f)) \\
   \varphi_2 &=& (f\geq g)  \land (g\geq h) \land (h\geq g) \\
   \varphi_3 &=& (f>g) \land \neg((h>g) \lor (f>h))
\end{eqnarray*}

The set of atoms $\mu_1=\set{(f>g), (f>h),(f=f),(g=g),(h=h)}$ is a
model for $\varphi_1$.  It satisfies the propositional part:
$\varphi_1$ evaluates to true when assigning the atoms in $\mu$ the
value ``true''. It satisfies the partial order part: it is a partial
order.
The set of atoms $\set{h>f,f>g}$ is not a model (for any partial order
constraint) because it is not closed under transitivity (nor
reflexivity). However, its extension
$\mu_2=\set{(h>f),(f>g),(h>g),(f=f),(g=g),(h=h)}$ is a model for
$\varphi_1$.
Formula $\varphi_1$ has additional models which are linearizations of
$\mu_1$: 
\[\begin{array}{l}
   \mu_3=\set{(f>g),(g>h),(f>h),(f=f),(g=g),(h=h)},\\
   \mu_4=\set{(f>h),(h>g),(f>g),(f=f),(g=g),(h=h)}, \text{ and}\\
   \mu_5=\set{(f>g),(g=h),(h=g),(f>h),(f=f),(g=g),(h=h)}
\end{array}\]
The formula $\varphi_2$ has a single model:
\[\set{(f=g),(g=f),(g=h),(h=g),(f=h),(h=f),(f=f),(g=g),(h=h)}\]

Focusing on $\varphi_3$ illustrates that there is an additional
implicit condition for an assignment to satisfy a partial order
constraint. We recall that a partial order can always be extended to a
total order. The partial order $\mu=\set{f>g}$ satisfies the
propositional part of $\varphi_3$ and may appear at first sight to
satisfy also the partial order part (it is a partial order). However,
no extension of $\mu$ to a total order satisfies the propositional
part of $\varphi_3$ and hence $\mu$ cannot be considered a model of
$\varphi_3$.
\end{example}

The following definition formalizes the semantics for partial order
constraints.

\begin{definition}[assignment, model]
An assignment $\mu$ is a mapping from propositions of $Atom_\cF$ to
truth values, and can be identified with the set of propositions it
assigns ``true''.
Let $\varphi$ be a partial order constraint on $\cF$. We say that an
assignment $\mu$ is a model for $\varphi$ if: (1) it makes $\varphi$
true as a propositional formula; (2) it satisfies the axioms for
strict partial order and equality; and (3) it defines a total order on
$\cF$.
More specifically, an assignment $\mu$ is required
to satisfy (for all $f,g,h\in\cF$):
\[
\begin{array}{lll}
\text{reflexivity:}&    (f=f)\in\mu &               \\
\text{symmetry:}&       (f=g)\in\mu ~~\Rightarrow~~ (g=f)\in\mu & \\
\text{asymmetry:}&      \neg((f>g)\in\mu \land (g>f)\in\mu) &  \\
\text{transitivity:}&
     (f>g)\in\mu~~\land~~ (g>h)\in\mu~~\Rightarrow~~ (f>h)\in\mu ~~& 
                                                               \\
                    &   
     (f=g)\in\mu~~\land~~ (g=h)\in\mu~~\Rightarrow~~ (f=h)\in\mu ~~& 
                                                               \\
\text{identity:}&       
     (f>g)\in\mu ~~\land~~ (g=h)\in\mu~~\Rightarrow~~ (f>h)\in\mu &
                                                               \\
                &       
     (f=g)\in\mu ~~\land~~ (g>h)\in\mu~~\Rightarrow~~ (f>h)\in\mu &
                                          \\
\text{comparability:~~}&  
     (f> g)\in\mu ~~\lor~~ (g>f)\in\mu ~~\lor~~~ (f=g)\in\mu&
                                            \\
\end{array}
\]
\end{definition}

Given that each model of a partial order constraint is a total order, we
have that $\neg(f>g) \equiv (g>f)\lor(g=f)$ and that $\neg (f=g)
\equiv (f>g)\lor (g>f)$. Hence we may assume without loss of
generality that partial order constraints are negation free. For example,
the formula $\varphi_3$ from Example~\ref{ex:syntax} is equivalent to
$\varphi_3' = (f>g) \land (g\ge h) \land (h\ge f)$ which is clearly
unsatisfiable.

\paragraph{\bf Satisfiability:~}

In this paper we are concerned with the question of satisfiability of
partial order constraints: given a partial order constraint $\varphi$
does it have a model? Similarly to the general SAT problem, the
satisfiability of partial order constraints is NP-complete, and the
reduction from SAT is straightforward.

The atoms in a formula $\varphi$ induce a graph $G_\varphi$ on the
symbols in $\cF$ such that satisfiability of $\varphi$ is equivalent to
that of its individual (strongly connected) components. This graph
captures all possible cycles in the partial order and hence all
potential contradictions. The following definition is inspired by
\cite{KK04}.

\begin{definition}[domain graph]
Let $\varphi$ be a (negation free) partial order constraint on
$\cF$. The domain graph $G_\varphi=(V,E)$ is a directed graph with
vertices $V=\cF$ and edges $E=\sset{(f,g)}{\set{(f>g),(f=g),
(g=f)}\cap Atom(\varphi) \neq\emptyset}.$
\end{definition}

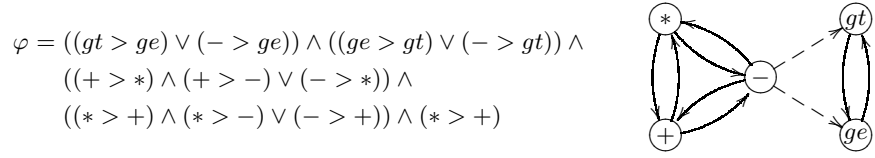
\begin{figure}[t]
\begin{minipage}{8.5cm}
\begin{eqnarray*}
\varphi &=& ((gt>ge)\lor (->ge)) \land ((ge>gt)\lor (->gt)) \land \\
        &&  ((+ > *)\land (+>-) \lor (-> *)) \land  \\
        && ((* > +)\land (* >-)\lor (-> +))\land (* > +)
\end{eqnarray*} 
\end{minipage}
\begin{minipage}{3.5cm}
\[
\xymatrix@R=10pt{
  \xyo{*}\ar@/_5pt/[dd]\ar@/_5pt/[dr]&&\xyo{gt}\ar@/_5pt/[dd]\\
  &\xyo{-}\ar@/_5pt/[ul]\ar@{-->}[ur]\ar@/_5pt/[dl]\ar@{-->}[dr]   &\\
  \xyo{+}\ar@/_5pt/[uu]\ar@/_5pt/[ur]&&\xyo{ge}\ar@/_5pt/[uu] }
\]
\end{minipage}
\caption{A partial order constraint (left) and its domain graph
  (right). The graph has two strongly connected components:
  $\{gt,ge\}$ and $\{-,*,+\}$. Arcs between the components are
  dashed.  }
\label{fig:graph}
\end{figure}

Figure~\ref{fig:graph} illustrates a partial order constraint~(a) and its
domain graph~(b). The following definition and lemma facilitate the
decomposition of a test for satisfiability to a set of smaller
instances.

\begin{definition}[restricting a partial order constraint]
Let $\varphi$ be a partial order constraint on $\cF$ and let
$F\subseteq\cF$.  The \emph{restriction} of $\varphi$ to the symbols
in $F$ is the formula obtained by substituting ``true'' for any atom
$(f\; R\; g)$ such that $(f,g)\not\in F\times F$.
The \emph{SCC-partition} of $\varphi$ is the set of graphs obtained by
restricting $\varphi$ to the nodes in each of the strongly connected
components of $G_\varphi$.
\end{definition}

\begin{example}\label{ex:scc-partition}
Consider the partial order constraint $\varphi$ 
and its domain graph $G_\varphi$ depicted as
Figure~\ref{fig:graph}. The graph $G_\varphi$ has two strongly
connected components. The SCC-partition for $\varphi$ gives:
\begin{eqnarray*}
\varphi_1 &=& ((gt>ge) \lor true) \land ((ge>gt) \lor true)
          \quad\equiv\quad true \\
\varphi_2 &=& (* > +) \land (((+ > *) \land (+ > -)) 
               \lor (- > *))  \land (((* > +) \land\\
          && (* > -)) \lor (- > +)) \quad
          \equiv \quad (* > +)\land (- > *)\land (- > +)
\end{eqnarray*}
\end{example}

\begin{lemma}\label{lemma:scc}
A partial order constraint is satisfiable if and only
if each of the formula in its SCC-partition is satisfiable.
\end{lemma}

\begin{proof} 
You can only get a contradiction if $x>x$ along some path in the
graph. Any such path will be contained in a single SCC.  
\end{proof}

\section{LPO  termination}
\label{sec:lpo}

A term rewrite system is a set of rules of the form $\ell\rightarrow
r$ where $\ell$ and $r$ are terms constructed from given sets of
symbols $\cF$ and variables $\cV$, and such that $r$ only contains
variables also in $\ell$. A rule $\ell\rightarrow r$ applies to a term
$t$ if a subterm $s$ of $t$ matches $\ell$ with some substitution
$\sigma$ (namely, $s=\ell\sigma$). The rule is applied by replacing
the subterm $s$ by $r\sigma$. Such an application is called a rewrite
step on $t$. A derivation is a sequence of rewrite steps.  A term
rewrite system is said to be terminating if all of its derivations are
finite. An example term rewrite system is depicted as
Figure~\ref{fig:trs}.

\begin{figure}[t]
\begin{centering}
\begin{minipage}{6cm}
\begin{SProg}
    $-gt(A,B) \rightarrow ge(B,A)$\\
    $-ge(A,B) \rightarrow gt(B,A)$ \\
    $-(A+B)     \rightarrow (-A)*(-B)$ \\
\end{SProg}
\end{minipage}
\begin{minipage}{6cm}
\begin{SProg}
    $-(A*B)     \rightarrow (-A)+(-B)$ \\
    $A*(A+B)    \rightarrow (A*B)+(A*C)$ \\
    $(B+C)*A    \rightarrow (B*A)+(C*A)$ \\
\end{SProg}
\end{minipage}
\end{centering}
\caption{An example term rewrite system: normalizing  formul\ae\/ with
  propositional connectives: $*$,$+$,$-$ (representing: and, or, not);
  and partial orders: gt, ge (representing: $>$,$\ge$).}
\label{fig:trs}
\end{figure}

Termination of term rewrite systems is undecidable. However a term
rewrite system terminates if there is a reduction ordering $\succ$ such
that $\ell\succ r$ for each rule $\ell\rightarrow r$ in the system. There
are many methods for defining such orderings. Many of them are based on
so-called simplification orderings and one such ordering is the
lexicographic path ordering (LPO)\cite{KL80,Der87}. 
In this setting a partial order $>_\cF$ (strict or non-strict) on
$\cF$ induces a corresponding (strict) partial order $>_{lpo}$ on
terms. If for each of the rules $\ell\rightarrow r$ in a system,
$\ell>_{lpo} r$ then the system is LPO terminating.

\begin{definition}[lpo]\label{def:lpo}
Let $>_\cF$ be a partial order (strict or non-strict) on $\cF$. The
induced lexicographic path ordering on terms constructed from
symbols of $\cF$ and variables from $\cV$ is defined recursively as
follows:
\begin{quote}
$s=f(s_1,\ldots,s_n) >_{lpo} t$ if and only if either:
\begin{enumerate}
\item
     $t=g(t_1,\ldots,t_m)$ and $s>_{lpo} t_j$, for all
     $1\le j\le m$, and either\\
     (a) $f>_\cF g$, or (b)  $f=g$ and 
     $\tuple{s_1,\ldots,s_n} >_{lpo}^{lex}\tuple{t_1,\ldots,t_m}$; or
\item
     $\fbox{$(s_i\approx t)$} \lor (s_i >_{lpo} t)$ for some $1\le i\le n$.
\end{enumerate}

\end{quote}
where $>_{lpo}^{lex}$ is the lexicographic
order on tuples of terms with respect to the ordering $>_{lpo}$:
\begin{eqnarray*}
\lefteqn{\tuple{s_1,\ldots,s_n} >_{lpo}^{lex} \tuple{t_1,\ldots,t_m}
\Leftrightarrow ~~ n>0 ~\land}\hspace{1cm}\\
&&  \left( m=0 ~\lor~
    m>0 ~\land 
    \left( \begin{array}{l} s_1>_{lpo} t_1 \quad \lor \\
            (\fbox{$s_1\approx t_1$} \land 
                           \tuple{s_2,\ldots,s_n} >_{lpo}^{lex} 
             \tuple{t_2,\ldots,t_m})
           \end{array}
    \right)\right)
\end{eqnarray*}
\end{definition}

The LPO termination problem is to determine for a given term rewrite
system $\cT$ with function symbols $\cF$, if there exists a partial
order $>_\cF$ such that $\ell >_{lpo} r$ for each of the rules
$\ell\rightarrow r\in\cT$ with the induced lexicographic path
ordering.  There are two variants of the problem: ``strict-'' and
``quasi-LPO termination'' depending on if we require $>_\cF$ to be
strict or not.
The corresponding decision problems, strict- and quasi- LPO
termination, are decidable and NP complete~\cite{Nieuwenhuis93}. These
problems are tantamount to testing corresponding partial order
constraints for satisfiability \cite{GieslTSF04,HirokawaM05}.
In Definition~\ref{def:lpo}, if $>_\cF$ is a strict order then the
boxed equivalences $s_i\approx t$ and $s_1\approx t_1$ are 
syntactic identity; and if $>_\cF$ is non-strict, then they are
an equivalence on terms: two terms are equivalent if they are
the same up to equivalent function symbols.

For given terms $s$ and $t$ the recursion in Definition~\ref{def:lpo}
for $s>_{lpo} t$ unfolds to a partial order constraint on the symbols
in $s$ and $t$, with details depending on whether $>_\cF$ is a strict
or non-strict partial order.
The strict- and quasi-LPO termination problems are to deciding if
conjunctions of these unfoldings are satisfiable --- one conjunct for
each rule in the given term rewrite system.

\begin{example}\label{ex:poftrs}
Consider the term rewrite system of Figure~\ref{fig:trs}. Unfolding
Definition~\ref{def:lpo} for strict-LPO termination, we obtain the
following:
\begin{small}
\begin{eqnarray*}
  -(gt(A,B)) >_{lpo} ge(B,A)         
        &\Longleftrightarrow& (gt>ge)\lor (->ge)\\
  -(ge(A,B)) >_{lpo} gt(B,A)         
      &\Longleftrightarrow& (ge>gt)\lor (->gt) \\
  -(A+B)     >_{lpo} (-(A)) * (-(B)) 
      &\Longleftrightarrow& (+ > *)\land ((+ > -)\land (+ > -))\lor (- > *) \\
  -(A*B)     >_{lpo} (-(A)) + (-(B)) 
      &\Longleftrightarrow& (* > +)\land ((* > -)\land (* > -))\lor (- > +)\\  
  A*(B+C)    >_{lpo} (A*B)+(A*C)     &\Longleftrightarrow& * > + \\
  (B+C)*A    >_{lpo} (B*A)+(C*A)     &\Longleftrightarrow& * > +
\end{eqnarray*}
\end{small}
The term rewrite system is LPO terminating if and only if the
conjunction of the constraints on the right sides is satisfiable. This
conjunction is precisely the partial order constraint $\varphi$ from
Figure~\ref{fig:graph} which by Lemma~\ref{lemma:scc}, $\varphi$ is
satisfiable if and only if the formula in its SCC-partition are.
Coming back to Example~\ref{ex:scc-partition}, it is straightforward
to observe that they are.
\end{example}

The next example illustrates a term rewrite system which is quasi-LPO
terminating but not strict-LPO terminating.

\begin{example}\label{sec:idiv}
Consider the following term rewrite system.

\medskip
\begin{SProg}
  $div(X,e) \rightarrow i(X)$\\
  $i(div(X,Y)) \rightarrow div(Y,X)$\\
  $div(div(X,Y),Z) \rightarrow div(Y,div(i(X),Z))$\\
\end{SProg}
\medskip

\noindent
Unfolding Definition~\ref{def:lpo} for strict-LPO gives
\begin{small}
\begin{eqnarray*}
   div(X,e) >_{lpo} i(X)   &\Longleftrightarrow& div>i   \\
   i(div(X,Y)) >_{lpo} div(Y,X)   &\Longleftrightarrow& i>div \\
   div(div(X,Y),Z) >_{lpo} div(Y,div(i(X),Z))   &\Longleftrightarrow& div>i
\end{eqnarray*}
\end{small}
The conjunction of the constraints on the right sides is not
satisfiable indicating that there does not 
exist any strict partial order on $\cF$ such that the corresponding
lexicographic path order decreases on the three rules.
The system is however quasi-LPO terminating. Unfolding
Definition~\ref{def:lpo} for quasi-LPO gives a satisfiable
partial order constraint equivalent to $(div\ge i) \wedge (i\ge div)$.
which indicates that taking $div=i$ provides a proof of quasi-LPO
termination.
\end{example}

\section{An Atom-based propositional encoding}\label{sec:Pencode}

The basic strategy is to encode a partial order constraint $\varphi$
on $\cF$ by an equivalent propositional formula $\varphi'$ such that
each model of $\varphi$ corresponds to a model of $\varphi'$ and in
particular such that $\varphi$ is satisfiable if and only if
$\varphi'$ is.
The idea of mapping strict LPO termination problems to a corresponding
propositional formula is addressed also in~\cite{KK04} where the authors
assume that partial order constraints contain only disjunction and
conjunction of atoms of the form $(f>g)$ (no equality and no negation).
This suffices for strict-LPO termination analysis. The presentation in
this section is more general and can be applied also for quasi-LPO
termination.

An atom-based propositional encoding for a partial order constraint
$\varphi$ is obtained by: (a) viewing the atoms in $\varphi$ as
propositional variables, and (b) making the axioms for partial order
explicit.
We let $\prop{a}$ denote the propositional variable corresponding to an
atom $a\in Atom_\cF$ and $\prop{\varphi}$ denote the propositional
formula obtained by replacing each atom $a$ in partial order constraint
$\varphi$ by the propositional variable $\prop{a}$.
For a set of symbols $\cF$ the following propositional formul\ae\/ make
the axioms explicit:

\medskip\noindent
\begin{minipage}{6cm}\small
\begin{itemize}
\item 
$R^=_\cF =
   \bigwedge\limits_{\scriptsize f \in\cF}
         \prop{f=f} $

\item
$A^>_\cF =
    \bigwedge\limits_{\scriptsize f,g\in\cF}
        \neg(\prop{f>g} \land \prop{g>f})$
\\
\item
$T^=_\cF =   \hspace{-9mm}
   \bigwedge\limits_{\scriptsize\begin{array}{c}f,g,h\in\cF\\
                           f\neq g\neq h\neq f\end{array}}   \hspace{-5mm}
        \prop{f=g} \land \prop{g=h} \rightarrow \prop{f=h}$

\item 
$I^2_\cF =   \hspace{-8mm}
   \bigwedge\limits_{\scriptsize\begin{array}{c}f,g,h\in\cF\\
                           f\neq g\neq h\neq f\end{array}}   \hspace{-5mm}
        \prop{f=g} \land \prop{g>h} \rightarrow \prop{f>h}$

\end{itemize}
\end{minipage}
\begin{minipage}{6cm}\small
\begin{itemize}
\item
$S^=_\cF=  
    \bigwedge\limits_{\scriptsize f,g\in\cF}
        \prop{f=g} \rightarrow \prop{g=f}$

\item 
$T^>_\cF =   \hspace{-9mm}
   \bigwedge\limits_{\scriptsize\begin{array}{c}f,g,h\in\cF\\
                           f\neq g\neq h\neq f\end{array}}   \hspace{-5mm}
        \prop{f>g} \land \prop{g>h} \rightarrow \prop{f>h}$

\item 
$I^1_\cF =   \hspace{-8mm}
   \bigwedge\limits_{\scriptsize\begin{array}{c}f,g,h\in\cF\\
                           f\neq g\neq h\end{array}}   \hspace{-5mm}
        \prop{f>g} \land \prop{g=h} \rightarrow \prop{f>h}$
\item 
$C^\ge_\cF =   \hspace{-8mm}
   \bigwedge\limits_{\scriptsize\begin{array}{c}f,g\in\cF\\
                           f\neq g\neq h\neq f\end{array}}   \hspace{-5mm}
        \prop{f>g} \lor \prop{g>f} \lor \prop{f=g}$
\end{itemize}
\end{minipage}
\medskip

The atom-based propositional encoding of a partial order constraint
$\varphi$ on symbols $\cF$ which does not involve equality nor
negation is obtained as $encode(\varphi)=\prop{\varphi} \land T^>_\cF
\land A^>_\cF$ \cite{KK04}.
In the general case when $\varphi$ may contain also negation or
equality the encoding is obtained as 
\begin{equation}\label{propositional}
encode(\varphi)=\prop{\varphi}
\land R^=_\cF\land S^=_\cF\land A^=_\cF \land T^>_\cF \land T^=_\cF
\land A^>_\cF \land I^1_\cF \land I^2_\cF\land C^\ge_\cF
\end{equation}

\begin{theorem}
A partial order constraint $\varphi$ on symbols $\cF$ is satisfiable
if and only if its atom-based propositional encoding $encode(\varphi)$
is.
\end{theorem}
\begin{proof}
  Straightforward.
\end{proof}

The two variants of atom-based propositional encodings both result in
large propositional formula. For the case when $|\cF|=n$ they
introduce $O(n^2)$ propositional variables and involve $O(n^3)$
clauses (e.g., for transitivity).

In \cite{KK04} Kurihara and Kondo propose two optimizations. They note
that for a given formula $\varphi$, the domain graph $G_\varphi$ is
often sparse and hence they propose to specialize the explicit
representation of the axioms for those symbols from $\cF$ actually
occurring in $\varphi$.
However, in view of Lemma~\ref{lemma:scc} we may assume that we are
testing satisfiability for partial order constraints which have
strongly connected domain graphs. Moreover, as indicated by our
experimental evaluation in Section~\ref{sec:implementation}, the
domain graphs for some of the more challenging examples strongly
connected components with up to 30 symbols.
Under this assumption, the proposed optimization will still require
$O(n^2)$ propositional variables and $O(n^3)$ clauses.

In a second optimization Kurihara and Kondo observe that the axioms
for transitivity and asymmetry can be replaced by a simpler axiom
(they call it $A^*$) introducing a single clause of the form
$\neg((f_1>f_2)\wedge(f_2>f_3) \wedge\cdots\wedge
(f_{k-1}>f_k)\wedge(f_k>f_1)$ for each simple cycle
$(f_1>f_2),(f_2>f_3),\ldots,(f_{k-1}>f_k),(f_k>f_1)$ in $G_\varphi$ to
assert that that cycle is not present in a model. They claim
correctness of the encoding and report considerable speedups when it
is applied.
The problem with this optimization is that in general there may be an
exponential number of simple cycles to consider.

The atom-based encoding described in this section either requires
$O(n^2)$ propositional variables and introduces $O(n^3)$ clauses or
else relies on a potentially exponential phase of processing the
simple loops in the domain graph.

\section{A Symbol-based propositional encoding}\label{sec:Sencode}

This section presents an alternative propositional encoding which
follows the same general strategy as before: It encodes a partial order
constraint by an equivalent propositional formula which in particular has
the same satisfiability. 
The novelty is in symbol-based approach. The basic idea is to
interpret the $n$ symbols in $\cF$ as indices in a partial order
taking finite domain values from the set $\set{1,\ldots,n}$. Each
symbol is thus modeled using $k=\lceil\log_2 n\rceil$ propositional
variables which encode the binary representation of its value.
Constraints of the form $(f>g)$ or $(f=g)$ on $\cF$ are interpreted as
constraints on indices and it is straightforward to encode them in
$k$-bit arithmetic.

Let $\varphi$ be a partial order constraint on $\cF$ and let $|\cF|=n$.
An integer \emph{solution} of $\varphi$ is an assignment $\theta$ of
the symbols in $\cF$ to values in $\{1\ldots,n\}$ which makes
$\varphi$ true.

\begin{example} 
  Consider again the partial order constraints from
  Example~\ref{ex:syntax}.  The assignments mapping $\tuple{f,g,h}$ to
  $\tuple{3,1,2}$, $\tuple{3,1,1}$ and $\tuple{1,1,1}$ are solutions for
  $\varphi_2$. But only the first is a solution for $\varphi_1$. The
  formula $\varphi_3$ has no solutions.
\end{example}

In the index based approach the semantics of a partial order
constraint is a set of integer solutions.

\begin{lemma}\label{lemma:s2m}
Let $\theta$ be a solution of $\varphi$. The assignment
\[
\mu = \sset{(f\; R\; g)}{ \{f,g\} \in \cF, ~R\in\cR,~ 
                                    (\theta(f)\; R\;\; \theta(g))}
\]
is a model of $\varphi$.
\end{lemma}

\begin{proof}
Clearly $\mu$ satisfies both the propositional and partial order parts
of $\varphi$ since the integer relation $>$ is a total order.  Hence
$\mu$ is a model for $\varphi$ by definition.
\end{proof}

\begin{lemma}\label{lemma:m2s}
Let $\mu$ be a model of $\varphi$ on $\cF$ with $n$ symbols. Then
there exists a solution $\theta$ of $\varphi$ in $\set{1,\ldots,n}$.
\end{lemma}

\begin{proof}
Assume $\cF = \{f_1, \ldots, f_n\}$ and let $\mu$ be a model of
$\varphi$.  
By asymmetry, identity and comparability, 
for each $1 \leq i < j \leq n$ exactly
one of $f_i > f_j$ or $f_i = f_j$ or $f_j > f_i$ hold.  We can
linearize the symbols in $\cF$: 
$f_{k_n}~ R_{n-1}~ \cdots R_{2}~ f_{k_2} ~R_{1}~ f_{k_1}$
where for each $1 \leq i < n$, $(f_{k_{i+1}}~ R_i~ f_{k_{i}})\in\mu$
and $R_i \in \{>, =\}$.
Since $\mu$ models transitivity, symmetry, and
identity.  We can then construct a solution $\theta$, using values
from $1$ to no more than $n$, where 
%
%
\[
\begin{array}{ll}
\theta(f_{k_1}) = 1 & \\ 
\theta(f_{k_{j+1}}) = \left\{ \begin{array}{ll}
    \theta(f_{k_j}) & \text{where~} R_{j-1} \equiv (=) \\
    \theta(f_{k_j})+1 & \text{where~} R_{j-1} \equiv (>) \\
  \end{array}
  \right. & \text{for~} 1\leq j < n
\end{array}
\]
\end{proof}

Note that there can be more than one solution corresponding to the
same model of $\varphi$ depending on how the symbols in $\cF$ are
linearized. The model $\mu_5 = \{f > g, f > h, g = h, h = g,
f=f,g=g,h=h\}$ corresponds to both $\theta_1 = \{f \mapsto 2,
g \mapsto 1, h \mapsto 1\}$ and $\theta_2 = \{f \mapsto 3, g \mapsto
2, h \mapsto 2\}$.

The following theorem is a direct consequence of
Lemmata~\ref{lemma:s2m} and~\ref{lemma:m2s}.

\begin{theorem}
A partial order constraint $\varphi$ has a solution iff it has a model.
\end{theorem}

We now introduce a symbol-based propositional encoding for partial
order constraints.  For  $|\cF|=n$ we need $k=\lceil\log
n\rceil$ bits per symbol. Recall that $\prop{a}$ is the propositional
variable corresponding to an atom $a$ and $\prop{\varphi}$ the
propositional formula obtained when replacing atoms by propositional
variables in partial order constraint $\varphi$.
\begin{enumerate}
\item For $f\in\cF$, the $k$-bit representation is
      $f=\tuple{f_k,\ldots,f_1}$ with $f_k$ the most significant
      bit. 
\item A constraint of the form $(f=g)$ is encoded in $k$-bits by
      \[ \encode{(f=g)}_k = \bigwedge_{i=1}^k (f_i\leftrightarrow
      g_i). \]
      A constraint of the form $(f>g)$ is encoded in $k$-bits by
        \[
          \encode{(f>g)}_{k} \Leftrightarrow 
             \begin{cases}   
              (f_1 \land \neg g_1) & k=1 \\
              (f_k \land \neg g_k) \lor 
              ((f_k \leftrightarrow g_k) \land \encode{(f>g)}_{k-1}) & k>1
              \end{cases}
        \]
\item 
      A partial order constraint $\varphi$ is encoded
      in $k$ bits by
        \begin{equation}\label{index}
           \encode{\varphi}_k = \prop{\varphi} \wedge
              \bigwedge_{a\in Atoms(\varphi)} 
                  (\prop{a} \leftrightarrow \encode{a}_k)
        \end{equation}
\end{enumerate}

\begin{proposition}
The size of the constraint based encoding of $\varphi$
is $O( |\varphi| \log n)$.
\end{proposition}

Let us compare the two encodings of a partial order constraint $\varphi$
given as Equations~(\ref{propositional}) and~(\ref{index}). The common
part in both encodings is the subformula $\prop{\varphi}$ in which
atoms are viewed as propositional variables. The difference is that
Equation~(\ref{propositional}) introduces explicit axioms to relate the
atoms in a partial order where Equation~(\ref{index}) interprets the
$n$ symbols as indices represented in $\lceil\log_2 n\rceil$-bits.
The symbol-based encoding introduces $O(n \log n)$ propositional
variables and involves $O(s \log n)$ clauses where $s$ is the size of
$\varphi$.  In comparison the atom-based encoding introduces
$O(n^2)$ variables and results in a formula with $O(n^3)$ clauses.
The key improvement is that the expensive encoding of the axioms is
not required because the encoding as integers ensures that they hold
``for free''.

\section{Implementation and Experimentation}
\label{sec:implementation}

We have implemented a prototype analyzer, poSAT, for strict- and
quasi- LPO termination based on the encoding proposed in
Section~\ref{sec:Sencode}.  The implementation is a written primarily
in SWI-Prolog \cite{swi-prolog,swi-page} and interfaces the MiniSat
solver \cite{EenS03,minisat-page} for solving SAT instances.
We have integrated MiniSat and SWI-Prolog through $\approx$190 lines of
C-code and $\approx$140 lines of Prolog code.  

\ignore{The C-code handles the low-level interface and conversion
  between the internal data representations of SWI-Prolog and of
  MiniSat. The Prolog code provides high-level interface for using the
  SAT solver in Prolog applications.}

The rest of poSAT is implemented in $\approx$700 lines of Prolog code.
This includes a TRS parser, modules to translate strict- and quasi-
LPO termination problems into partial order constraints, the module
converting partial order constraints into SAT instances, and finally a
head module processing the command line, running the components,
pretty-printing the results etc.
The current implementation does not decompose partial order
constraints to their SCC-components (Lemma~\ref{lemma:scc}). The
experimental results indicate that the implementation would not
benefit from that: (a) Most of the tests are very fast without this
decomposition; and (b) It is typical for hard cases of LPO termination
(see Table~\ref{tab:hardest}) to have a large strongly connected
component including the majority of the symbols.

For experimentation we have taken all 751 term rewrite systems from the
Termination Problem Data Base \cite{tpdb} which do not specify a
``theory'' or a ``strategy''. In the following, the names of term
rewrite systems are indicated in typewriter font and can be found in
\cite{tpdb}.
We report on the comparison of poSAT for both strict- and quasi-LPO
termination analysis with the TTT analyzer\cite{ttt}. We have also
performed comparisons with AProVe \cite{aprove} as well as with the
results reported in \cite{KK04}.

For the experiments, poSAT runs on a 1.5GHz laptop running GNU/Linux
FC4.
The TTT analyzer is applied via its Web interface \cite{ttt} and runs on
a Xeon 2.24GHz dual-CPU platform which is a considerably faster machine
than ours. 
%
%
Experiments with AProVe running on our local platform give results
which are consistently slower than TTT (on its faster machine). Hence
for comparison with poSAT we provide the numbers only for TTT.
Comparison with the results of \cite{KK04} are also not presented as
they too are substantially slower than those obtained with TTT.

With regards to precision, as expected, all three analyzers give the
same results (with the exception of a single test which TTT cannot
handle within the maximum timeout allocation). From the 751 example
systems, 128 are LPO terminating and 132 are quasi LPO terminating.
For poSAT, run times include the complete cycle of processing each test:
reading and parsing the file, translation to partial order constraints and
then to propositional formula, solving by the SAT solver and printing
the results. The run time of each test is computed as an average of ten
identical runs.

\begin{table}[t]

\quad\quad\begin{minipage}{6cm} 
\begin{tabular}{|r|r|r|}
  \hline
  {}      & \multicolumn{1}{c|}{\:\:poSAT\:\:} & \multicolumn{1}{c|}{TTT} \\
  \hline                  
  Total   & 9.112 & 302.35\\
  Average & 0.012 &  0.40  \\
  Max     & 0.450 & 254.47\\
  \hline
  \multicolumn{3}{c}{}\\
  \multicolumn{3}{c}{{\bf (a)} strict LPO termination}\\[1ex]
\end{tabular}
\end{minipage}
\begin{minipage}{6cm}
\begin{tabular}{|r|r|r|}
  \hline
  {}      & \multicolumn{1}{c|}{\:\:poSAT\:\:} & \multicolumn{1}{c|}{TTT} \\
  \hline
  Total   & 10.428 & 2167.44   \\
  Average &  0.014 &   2.89    \\
  Max     &  1.169 &  600.00  \\
  \hline 
  \multicolumn{3}{c}{}\\
  \multicolumn{3}{c}{{\bf (b)} quasi LPO termination}\\[1ex]
\end{tabular}
\end{minipage}

\caption{Summary of experimental results: total, average and maximum
  times (sec) for 751 tests.} \label{table:summary}

\end{table}

Table~\ref{table:summary}(a) summarizes the results for strict LPO
termination analysis. The columns contain times (in seconds) for our
analyzer ({\bf poSAT}) and TTT. We configure TTT to run with timeout of
10 minutes, the maximum allowed by its Web interface.
Note that the times are taken on different machines which makes the
precise comparison impossible. Nevertheless, the results are indicative
showing that poSAT is fast in absolute terms and scales better for hard
cases. Notably, the hardest test of LPO termination for poSAT ({\tt
  HM/t005.trs}) completes in under a half second, while the hardest test
for TTT ({\tt currying/Ste92/hydra.trs}) takes more than 4 minutes.

Table~\ref{table:summary}(b) presents the results for quasi LPO
termination analysis. For this variant, poSAT completes the 751 tests in
10.43sec. The same task takes TTT over 36 \emph{minutes} with one test
({\tt currying/Ste92/hydra.trs}) running out of 10 minutes timeout. The
next hardest test for TTT is {\tt currying/AG01\_No\_3.13.trs} which
completes in 203.9sec (3.4min). The same two tests take poSAT 0.01sec
and 0.031sec respectively.
%
The hardest quasi LPO test for  poSAT's is {\tt Zantema/z30} which
takes 1.17sec in our analyzer and 5.03sec in TTT.

Once again, the timings are indicative despite the fact that the two
analyzers run on different machines. By comparing the results in
Table~\ref{table:summary}(a) and (b) we observe that for quasi LPO,
TTT runs about an order of magnitude slower than for strict LPO. In
contrast, poSAT demonstrates only a modest 14\% increase in the
accumulated run time.

\ignore{We have collected profiling results for the two whole runs of
  our analyzer in strict and non-strict LPO mode. The results in both
  cases indicate that each of the three stages namely, parsing of the
  TRS, translation to partial order and SAT formula, and solving the
  formula by MiniSat account roughly for 1/3 of the run time. \vit{It
    means that by further optimizing parsing and translation,
    re-implementing in C, etc.  we can bring down the total time}}

\begin{table}[t]
  
  \centering{\small
    \begin{tabular}{l|r|r|r|r|r|r|r|r}
      \multicolumn{1}{c|}{} &
      \multicolumn{4}{c|}{LPO} & \multicolumn{4}{c}{quasi-LPO}\\
      \hline
      \multicolumn{1}{c|}{Test} &
      \multicolumn{1}{c|}{Sym} &
      \multicolumn{1}{c|}{CNF} &
     \multicolumn{1}{c|}{poSAT} &
       \multicolumn{1}{c|}{TTT} &
      \multicolumn{1}{c|}{Sym} &
      \multicolumn{1}{c|}{CNF} &
      \multicolumn{1}{c|}{poSAT} &
      \multicolumn{1}{c}{TTT}
      \\
      \hline
      \hline
      { AProVE/AAECC-ring }            &28/10  &642/2479      &0.089      &0.04 &28/24  &786/3118      &0.110     &0.35\\
      { Cime/mucrl1               }    &0/0    &0/1           &0.294      &2.51 &0/0    &0/1           &0.236     &19.85\\
      { currying/AG01\_No\_3.13   }    &0/0    &0/1           &0.127      &0.01 &0/0    &0/1           &0.031     &203.92\\
      { currying/Ste92/hydra     }    &8/4    &102/337       &0.020      &254.5&8/6    &1156/3570     &0.103     &$\infty$\\
      { HM/t005                   }    &0/0    &0/1           &0.450      &0.05 &0/0    &0/1           &0.038     &2.22\\
      { HM/t009                   }    &19/11  &773/3139      &0.161      &0.00 &19/17  &1388/5677     &0.209     &0.15\\
      { /Ex1\_2\_AEL03\_C          }    &19/17  &630/2506      &0.113      &0.00 &19/19  &1286/5416     &0.169     &95.00\\
      { /Ex1\_2\_AEL03\_GM         }    &22/17  &506/1921      &0.062      &0.00 &22/22  &693/2707      &0.069     &19.37\\
      { /Ex26\_Luc03b\_C           }    &15/12  &384/1450      &0.061      &0.02 &15/15  &816/3260      &0.099     &6.21\\
      { /Ex2\_Luc02a\_C            }    &15/12  &390/1477      &0.062      &0.10 &15/15  &838/3360      &0.097     &6.03\\
      { /Ex4\_7\_37\_Bor03\_C      }    &13/11  &287/1057      &0.062      &0.04 &13/12  &577/2317      &0.081     &0.82\\
      { /Ex5\_7\_Luc97\_C          }    &18/15  &614/2413      &0.098      &0.01 &18/18  &1341/5533     &0.173     &94.90\\
      { /Ex5\_7\_Luc97\_GM         }    &22/19  &499/1876      &0.056      &0.01 &22/20  &752/2947      &0.074     &19.29\\
      { /Ex6\_15\_AEL02\_C         }    &23/22  &906/3658      &0.159      &0.01 &23/23  &1862/7717     &0.272     &126.02\\
      { /Ex6\_15\_AEL02\_FR        }    &26/20  &599/2284      &0.072      &0.01 &26/26  &867/3430      &0.080     &10.00\\ 
      { /Ex6\_15\_AEL02\_GM        }    &29/25  &745/2929      &0.085      &0.03 &29/29  &1074/4297     &0.105     &157.86\\
      { /Ex6\_15\_AEL02\_Z         }    &26/20  &587/2236      &0.061      &0.00 &26/26  &869/3460      &0.078     &18.24\\
      { /Ex7\_BLR02\_C             }    &14/11  &299/1108      &0.048      &0.05 &14/14  &627/2546      &0.073     &1.67\\
      { /Ex9\_BLR02\_C             }    &12/9   &296/1087      &0.054      &0.28 &12/10  &608/2390      &0.069     &0.35\\
      { /ExAppendixB\_AEL03\_C     }    &20/18  &700/2809      &0.134      &0.00 &20/20  &1410/5920     &0.212     &113.12\\
      { /ExIntrod\_GM99\_C         }    &16/13  &423/1591      &0.084      &0.00 &16/14  &848/3416      &0.105     &21.69\\
      { /ExIntrod\_Zan97\_C        }    &15/12  &344/1285      &0.057      &0.01 &15/15  &709/2862      &0.081     &2.04\\
      { /ExSec11\_1\_Luc02a\_C     }    &16/13  &439/1666      &0.069      &0.00 &16/16  &985/3914      &0.121     &29.38\\
      { Zantema/z01               }    &2/2    &84/247        &0.028      &0.01 &3/3    &2439/7315     &0.198     &0.12\\
      { Zantema/z30               }    &2/2    &65/190        &0.119      &0.01 &3/2    &12827/38479   &1.169     &5.03\\
\end{tabular}
}
\caption{The 25 hardest tests for poSAT}
\label{tab:hardest}
\end{table}

Table~\ref{tab:hardest} presents a detailed analysis for the 25 most
challenging examples for poSAT chosen by maximum total time for
strict- and quasi- LPO analysis. 
The two parts of the table present the respective results for strict- and
quasi-LPO termination analyses.
The following information is provided:
The columns labeled ``Sym'' and ``CNF'' characterize the partial order
constraints derived from the given term rewrite systems. ``Sym''
indicates the number of symbols in the complete formula and in the
largest component of its SCC-partition (0/0 in this column means that
the partial order constraint is trivial i.e., true or false). ``CNF''
indicates the numbers of propositional variables and clauses in the
translation of the propositional (symbol-based) encoding to
conjunctive normal form.
The columns labeled  ``poSAT'' and ``TTT'' indicate run times (in
seconds) for the poSAT and TTT solvers.

All of the tests except for {\tt currying/Ste92/hydra.trs} are not
strict- nor quasi-LPO terminating. This is not surprising for the 25
hardest tests, as proving unsatisfiability is harder than finding a
solution for a satisfiable formula.
It is interesting to note that three examples among the hardest 25,
result in trivial partial order constraints.  Obviously, the challenge
in these examples is not in solving the constraints but rather in
obtaining them by unfolding Definition~\ref{def:lpo}. 
Interestingly, our translation and simplification mechanisms are
sometimes more powerful than those of TTT.  For instance, {\tt
currying/AG01\_No\_3.13} is simplified to false in poSAT but not in
TTT, leading to a long search for TTT. The difference is due to the
fact that in the case of poSAT the generation of a partial order
formula never introduces trivial sub-formula (``true'' or ``false'').
these are evaluated on-the-fly.

Another observation based on the results of Table~\ref{tab:hardest} is
that the partial order constraints derived from the tests typically
have domain graphs with large strongly-connected components. Almost
every test in the table has a ``core'' component including the
majority of the symbols. Therefore, it is unlikely that the performance
of poSAT for the presented tests can be improved by using the
SCC-based decomposition of the formula.  Large components of this size
would also pose a serious challenge to an analyzer based on the
atom-based encoding.

As Table~\ref{tab:hardest} shows, the maximum CNF instance solved in our
tests includes 12827 propositional variables and 38479 CNF clauses. This
is well below the capacity limits of MiniSat, which is reported to
handle benchmarks with hundreds of thousands of variables and clauses
\cite{minisat-page}.

\section{Related and Future Works}

Testing for satisfiability of partial order constraints comes up in
many other applications. First of all in the context of term rewrite
systems where LPO is just one example of a simplification ordering and
analyses based on other types of orderings may also be encoded into
propositional logic.
Moreover, for programs which cannot be shown to terminate using these
kinds of simplification orderings, the dependency pairs approach
\cite{ArtsG97,ArtsG00} has proven very successful in generating sets
of constraints such that the existence of a (quasi-)ordering
satisfying them is a sufficient condition for termination. Our
constraint solving technique is directly applicable and will improve
considerably the performance of implementations for these techniques.

Another approach to proving termination is based on the notion of
\emph{size change graphs} \cite{Jones:POPL01} which approximate the
transition relation induced by a program.  A transition step from
program point $p/n$ (with $n$ state variables) to program point $q/m$
(with $m$ state variables) is described by a graph with nodes
$\set{p_1,\ldots,p_n}$ and $\set{q_1,\ldots,q_m}$ and an edge from
$p_i$ to $q_j$ to indicate that the size of the $i^{th}$ state
variable at point $p$ decreases (strict or non-strict) on the
transition with respect to the $j^{th}$ state variable at point $q$.
Size change graphs can be represented as conjunctions of constraints
of the form $(p_i>q_j)$ or $(p_i\ge q_j)$ (see for example
\cite{CLS:iclp2005}). Sets of size change graphs can be represented as
disjunctions of conjunctions of such constraints and encoded as
propositional formula using using $\log (n+m)$ bits. The results in
\cite{CLSS:esop2005} illustrate how such an encoding can be exploited
to provide for efficient implementations manipulating potentially
large sets of size change graphs.

\section{Conclusion}

We have introduced a new kind of propositional encoding for reasoning
about partial orders. Previous works propose to represent the atoms in
a formula as propositional variables and to explicitly encode
the axioms for partial order. Our novel approach is to interpret the
symbols in a formula as finite domain variables corresponding to the
indices in the partial order.
We illustrate the application of our approach for LPO termination
analysis for term rewrite systems. Experimental results are
unequivocal indicating orders of magnitude speedups in comparison with
current implementations for LPO termination analysis.
The proposed technique is directly applicable to more powerful
termination proving techniques, such as those based on dependency
pairs \cite{ArtsG00}, which basically involve the same kind of
constraint solving.

\subsection*{Acknowledgment}

We are grateful to Bart Demoen for useful insights regarding the
implementation and to Samir Genaim who donated the Prolog parser for
term rewrite systems.
J{\"u}rgen Giesl and Aart Middeldorp assisted with
the use of the AProVe and TTT analysers. Masahito Kurihara  provided
the test cases from \cite{KK04}.


\begin{thebibliography}{10}

\bibitem{ArtsG97}
T.~Arts and J.~Giesl.
\newblock Automatically proving termination where simplification orderings
  fail.
\newblock In M.~Bidoit and M.~Dauchet, editors, {\em TAPSOFT'97: Theory and
  Practice of Software Development, 7th International Joint Conference
  CAAP/FASE, Lille, France, April 14-18, 1997, Proceedings}, volume 1214 of
  {\em Lecture Notes in Computer Science}, pages 261--272. Springer, 1997.

\bibitem{ArtsG00}
T.~Arts and J.~Giesl.
\newblock Termination of term rewriting using dependency pairs.
\newblock {\em Theor. Comput. Sci.}, 236(1-2):133--178, 2000.

\bibitem{aprove}
Automated program verification environment (aprove).
\newblock http://www-i2.informatik.rwth-aachen.de/AProVE/.
\newblock Viewed December 2005.

\bibitem{CLSS:esop2005}
M.~Codish, V.~Lagoon, P.~Schachte, and P.~J. Stuckey.
\newblock Size-change termination analysis in $k$-bits.
\newblock In {\em Proceedings}, Lecture Notes in Computer Science. Springer,
  2005.

\bibitem{CLS:iclp2005}
M.~Codish, V.~Lagoon, and P.~J. Stuckey.
\newblock Testing for termination with monotonicity constraints.
\newblock In M.~Gabbrielli and G.~Gupta, editors, {\em Logic Programming, 21st
  International Conference, ICLP 2005, Sitges, Spain, October 2-5, 2005,
  Proceedings}, volume 3668 of {\em Lecture Notes in Computer Science}, pages
  326--340. Springer, 2005.

\bibitem{Der87}
N.~Dershowitz.
\newblock Termination of rewriting.
\newblock {\em J. Symb. Comput.}, 3(1/2):69--116, 1987.

\bibitem{DJ90}
N.~Dershowitz and J.-P. Jouannaud.
\newblock Rewrite systems.
\newblock In J.~van Leeuwen, editor, {\em Handbook of Theoretical Computer
  Science}, volume B: Formal Models and Semantics, pages 2435--320. Elsevier
  and MIT Press, 1990.

\bibitem{EenS03}
N.~E{\'e}n and N.~S{\"o}rensson.
\newblock An extensible sat-solver.
\newblock In E.~Giunchiglia and A.~Tacchella, editors, {\em Theory and
  Applications of Satisfiability Testing, 6th International Conference, SAT
  2003 (Selected Revised Papers)}, volume 2919 of {\em Lecture Notes in
  Computer Science}, pages 502--518. Springer, 2004.

\bibitem{GieslTSF04}
J.~Giesl, R.~Thiemann, P.~Schneider-Kamp, and S.~Falke.
\newblock Automated termination proofs with {AProVE}.
\newblock In V.~van Oostrom, editor, {\em Rewriting Techniques and
  Applications, 15th International Conference, RTA 2004, Aachen, Germany, June
  3-5, 2004, Proceedings}, volume 3091 of {\em Lecture Notes in Computer
  Science}, pages 210--220. Springer, 2004.

\bibitem{HirokawaM05}
N.~Hirokawa and A.~Middeldorp.
\newblock Tyrolean termination tool.
\newblock In {\em Term Rewriting and Applications, 16th International
  Conference, RTA 2005, Nara, Japan, April 19-21, 2005, Proceedings}, volume
  3467 of {\em Lecture Notes in Computer Science}, pages 175--184. Springer,
  2005.

\bibitem{KL80}
S.~Kamin and J.-J. Levy.
\newblock Two generalizations of the recursive path ordering.
\newblock Department of Computer Science, University of Illinois, Urbana, IL.
  Available at
  http://www.ens-lyon.fr/LIP/REWRITING/OLD\_PUBLICATIONS\_ON\_TERMINATION/
  (viewed December 2005), 1980.

\bibitem{KK04}
M.~Kurihara and H.~Kondo.
\newblock Efficient {BDD} encodings for partial order constraints with
  application to expert systems in software verification.
\newblock In {\em Innovations in Applied Artificial Intelligence, 17th
  International Conference on Industrial and Engineering Applications of
  Artificial Intelligence and Expert Systems, IEA/AIE 2004, Ottawa, Canada, May
  17-20, 2004. Proceedings}, volume 3029 of {\em Lecture Notes in Computer
  Science}, pages 827--837. Springer, 2004.

\bibitem{Jones:POPL01}
C.~S. Lee, N.~D. Jones, and A.~M. Ben-Amram.
\newblock The size-change principle for program termination.
\newblock {\em ACM SIG{\-}PLAN Notices}, 36(3):81--92, 2001.
\newblock Proceedings of POPL'01.

\bibitem{minisat-page}
Mini{SAT} solver.
\newblock http://www.cs.chalmers.se/Cs/Research/FormalMethods/MiniSat.
\newblock Viewed December 2005.

\bibitem{Nieuwenhuis93}
R.~Nieuwenhuis.
\newblock Simple {LPO} constraint solving methods.
\newblock {\em Inf. Process. Lett.}, 47(2):65--69, 1993.

\bibitem{swi-page}
Swi-prolog.
\newblock http://http://www.swi-prolog.org/.
\newblock Viewed December 2005.

\bibitem{tpdb}
The termination problems data base.
\newblock http://www.lri.fr/$\sim$marche/tpdb/.
\newblock Viewed December 2005.

\bibitem{ttt}
Tyrolean termination tool.
\newblock http://cl2-informatik.uibk.ac.at/ttt/.
\newblock Viewed December 2005.

\bibitem{swi-prolog}
J.~Wielemaker.
\newblock An overview of the {SWI-Prolog} programming environment.
\newblock In F.~Mesnard and A.~Serebenik, editors, {\em Proceedings of the 13th
  International Workshop on Logic Programming Environments}, pages 1--16,
  Heverlee, Belgium, Dec. 2003. Katholieke Universiteit Leuven.
\newblock CW 371.

\end{thebibliography}

\end{document}